\documentclass[11pt, oneside, letterpaper, fleqn, leqno]{article}

\usepackage[utf8]{inputenc}
\usepackage[T1]{fontenc}
\usepackage[top=1in, bottom=1in, left=1in, right=1in]{geometry}
\usepackage{fouriernc}
\usepackage{authblk}
\usepackage{mathtools}
\usepackage{multicol}
\usepackage{graphicx}
\usepackage{amssymb}
\usepackage{bbold}
\usepackage{appendix}

\newcommand{\pd}{\partial}

\title{Supersymmetry of the quantum rotor}

\author[$\dag$]{Vincent X. Genest}

\author[$\ddag$]{Luc Vinet}

\author[*]{Guo-Fu Yu}

\author[**]{Alexei Zhedanov}
\affil[$\dag$]{Department of Mathematics, Massachusetts Institute of Technology 

Cambridge MA 02139, USA}
\affil[$\ddag$]{Centre de recherches math\'ematiques, Universit\'e de Montr\'eal
	
Montr\'eal QC H3C 3J7, Canada}

\affil[*]{Department of Mathematics, Shanghai Jiao Tong University
	
Shanghai 200240, China}

\affil[**]{Department of Applied Mathematics and Physics, Graduate School of Informatics, Kyoto University, Kyoto 606-8501, Japan}

\date{}

\begin{document}
\maketitle
\thispagestyle{empty}
\hrule
\begin{abstract}
\noindent
The quantum rotor is shown to be supersymmetric. The supercharge $Q$, whose square equals the Hamiltonian, is constructed with reflection operators. The conserved quantities that commute with $Q$ form the algebra $so(3)_{-1}$, an anticommutator version of $so(3)$. The subduced representation of $so(3)_{-1}$ on the space of spherical harmonics with total angular momentum $j$ is constructed and found to decompose into two irreducible components. Two natural bases for the irreducible representation spaces of $so(3)_{-1}$ are introduced and their overlap coefficients prove expressible in terms of orthogonal polynomials of a discrete variable called anti-Krawtchouk polynomials.
\end{abstract}
\hrule
\begin{center}
\emph{This paper is cordially dedicated to Mourad Ismail on the occasion of his 70\textsuperscript{th} birthday.}
\end{center}
\section*{Introduction}
It has been appreciated that certain scalar quantum mechanical systems can exhibit supersymmetry with supercharges involving reflection operators or more generally involutions \cite{Plyuschay-1996, Post-2011}. The simplest example of such an instance is that of the following harmonic oscillator Hamiltonian in one dimension \cite{Post-2011}:
\begin{equation*}
H = - \frac{1}{2} \frac{\pd^2}{\pd x^2} 
+ \frac{1}{2} x^2
- \frac{1}{2} R,
\end{equation*}
where $R$ is the reflection operator acting as $x$: $R f(x) = f(-x)$. Indeed, it is straightforward to see that one then has
\begin{equation*}
H = Q^2,
\end{equation*}
with the supercharge $Q$ having the expression
\begin{equation*}
Q= \frac{1}{\sqrt{2}} \left( \frac{\pd}{\pd x}R + x \right).
\end{equation*}
The occurence of involutions in quantum mechanical dynamics is not unfamiliar \cite{Brink-1993, Genest-2014, Lapointe-1996, Poly-1992}; it is in particular related to the use of Dunkl operators which are differential-difference operators associated to reflection groups \cite{Dunkl-1989}. The purpose of this paper is to show that another simple system, the quantum linear rigid rotor, is similarly supersymmetric.

When vibrational modes are neglected, a diatomic molecule can be described by the quantum rotor Hamiltonian which is nothing else than the $so(3)$ Casimir operator
\begin{equation}
\label{H-Rigid}
H = J_1^2 + J_2^2 + J_3^2 + 1/4,
\end{equation}
where the $J_1$, $J_2$, $J_3$ are the angular momentum generators
\begin{equation*}
J_1 = \frac{1}{i} (x_2 \pd_{x_3} - x_3 \pd_{x_2}),
\quad
J_2 = \frac{1}{i} (x_3 \pd_{x_1} - x_1 \pd_{x_3}),
\quad 
J_3 = \frac{1}{i} (x_1 \pd_{x_2} - x_2 \pd_{x_1}).
\end{equation*}
We are choosing units in which $\hbar^2/2I = 1$, with $I$ the moment of inertia, and are shifting the energies by $1/4$ for convenience. It is well known from angular momentum theory that the generators $J_i$'s are symmetries of the quantum rotor, i.e.
\begin{equation*}
[H, J_1] = 0, \quad [H, J_2] = 0, \quad [H, J_3] = 0,
\end{equation*}
and that they satisfy the commutation relations
\begin{equation}
\label{so3-comm}
[J_1, J_2] = i J_3, \quad [J_2, J_3] = i J_1, \quad [J_3, J_1] = i J_2.
\end{equation}
The spectrum of $H$ is positive with the energies labeled by a non-negative integer $j$ and given by
\begin{equation*}
E_{j} = (j + 1/2)^2,\qquad j = 0, 1, 2,\ldots
\end{equation*}
The fact that $E_{j}$ is a square is an indication that the Hamiltonian might be supersymmetric. Owing to the $so(3)$ symmetry of $H$, the level $E_{j}$ has a $(2j + 1)$-fold degeneracy. A natural basis for the wavefunctions is obtained by simultaneously diagonalizing $J_3$ and $H$. This leads to the spherical harmonics $Y_{j}^{m}(\hat{r})$, where $\hat{r} \in S^2$, which satisfy \cite{Edmonds-1996} 
\begin{align}
\label{J3}
\begin{aligned}
&H \, Y_{j}^{m}(\hat{r}) = E_{j} Y_{j}^{m}(\hat{r}),\qquad j = 0, 1, 2, \ldots
\\
& J_3 \, Y_{j}^{m}(\hat{r}) = m Y_{j}^{m}(\hat{r}),\qquad m = -j, -j+1, \ldots, 0, \ldots,j - 1, j,
\end{aligned}
\end{align}
and which have the expression
\begin{equation}
\label{Yjm}
Y_{j}^{m} (\hat{r})= \sqrt{\frac{2j + 1}{4\pi} \frac{(j - m)!}{(j + m)!}}\; P_{j}^{m}\left(\frac{x_3}{\sqrt{x_1^2 + x_2^2 + x_3^2}}\right)\; \left(\frac{x_1 + i x_2}{\sqrt{x_1^2 + x_2^2}}\right)^{m},
\end{equation}
where $P_{j}^{m}(z)$ is the associated Legendre polynomials. They satisfy the orthogonality relation
\begin{equation*}
\int_{S^2} Y_{j}^{m}(\hat{r})\,(Y_{j'}^{m'}(\hat{r}))^*\;\mathrm{d}\Omega = \delta_{j j'} \delta_{m m'},
\end{equation*}
where $a^*$ stands for the complex conjugate of $a$. For a fixed value of $j$, the spherical harmonics $Y_{j}^{m}(\hat{r})$ span an irreducible representation space of $so(3)$ with the raising and lowering operators $J_{\pm} = J_1 \pm i J_2$ acting as follows:
\begin{equation}
\label{Jpm-Action}
J_{\pm}\,Y_{j}^{m}(\hat{r}) = [(j \mp m)(j \pm m + 1)]^{1/2} Y_{j}^{m \pm 1}(\hat{r}).
\end{equation}
Let $R_1$, $R_2$, $R_3$ be the operators that respectively reflect the coordinates $x_1$, $x_2$, $x_3$. With $\{A, B\}= AB + BA$, it is easily verified that
\begin{align}
\label{r-comm}
\begin{aligned}
&\{L_i, R_j\} = 0, \qquad i\neq j,
\\
&[L_i, R_i] = 0.
\end{aligned}
\end{align}
It follows that $H$ is also invariant under all reflections $R_i$, $i = 1,2,3$. The constants of the motion of $H$ are thus elements of the universal enveloping algebra of $so(3)$ combined multiplicatively with reflections. This set can be viewed as spanning the full symmetry algebra of the quantum rotor. Within this ensemble, there are elements $Q$ whose square is equal to $H$; this explains why we claim that $H$ is supersymmetric.

Given one such $Q$, we shall consider the subset of conserved quantities that have the property of commuting with $Q$ and shall identify its structure. It will be shown to be a unital algebra freely generated by operators obeying relations expressed as the anticommutator analogs of the $so(3)$ relations \eqref{so3-comm}. This algebra will be denoted $so(3)_{-1}$ since, as shall be explained, it is the $q = -1$ version of $so(3)_{q}$, a $q$-deformation of $so(3)$ that has been introduced and studied in \cite{Fairlie-1990, Havlicek-1999}. The algebra $so(3)_{-1}$ has already been considered per se and its representations have been examined \cite{Arik-2003,Brown-2013, Genest-2014-2, Gorodnii-1984, Ostrovskyi-1992}. It is sometimes referred to as the anticommutator spin algebra. We shall here construct its representations on the space of wavefunctions of the rotor, i.e. on the spherical harmonics. These representations are subduced from the $(2j + 1)$-dimensional irreducible representations of the full symmetry algebra $U(so(3))\otimes \mathbb{Z}_2^3$ of $H$. For a given value of $j$, these $so(3)_{-1}$ representations will be found to have two irreducible components of dimension $j + 1$ and $j$.

The idea of an operator whose square is equal to a Laplacian is familiar in spinorial contexts where $\gamma$-matrices or Clifford algebra generators are used. The fact that this can be achieved in a scalar situation using reflections/involutions is much less known. This was observed in the framework of two of the author's analysis of the Laplace--Dunkl equation on $S^2$ \cite{Genest-2015}. It is also connected to the tensoring of superalgebra realizations when the grade involution is employed in a manifest fashion \cite{Genest-2016}. In light of these observations much of Clifford analysis can be translated to the scalar realm. Our goal here is to indicate how this pans out in a simple and physically relevant model.

The remainder of the paper proceeds as follows. In Section 1, we introduce the supercharge $Q$ and exhibit the generators of its commutant. We identify the structure relations they satisfy and indicate why the notation $so(3)_{-1}$ is appropriate for the algebra they generate. The non-uniqueness of the supercharge will also be commented upon. In Section 2, we construct the representations of $so(3)_{-1}$ on the space of spherical harmonics $Y_{j}^{m}(\hat{r})$ which will be seen to split in 2 irreducible $so(3)_{-1}$ modules. In Section 3, we shall indicate how the so-called anti-Krawtchouk polynomials define the transformation matrices between two natural bases for the irreducible representations of $so(3)_{-1}$.
\section{The supercharge $Q$ and its symmetry algebra}
In this Section, we exhibit the supercharge $Q$ which squares to the Hamiltonian \eqref{H-Rigid} of the quantum rigid rotor. We obtain its symmetries, and show that they generate $so(3)_{-1}$, also known as the anticommutator spin algebra.

Consider the operator $Q$ defined as
\begin{equation}
\label{Q-op}
Q = -i J_1 R_3 + i J_2 R_2 R_3 - i J_3 R_2 -1/2.
\end{equation}
Upon using the relations \eqref{so3-comm} and \eqref{r-comm}, a straightforward calculation shows that one has
\begin{equation}
\label{Main-rel}
Q^2 = H,
\end{equation}
where $H$ is the Hamiltonian of the rigid quantum rotor given by \eqref{H-Rigid}. Since the reflection operators $R_i$ are self-adjoint, that is $R_i^{\dagger} = R_i$, it can be verified using \eqref{r-comm} that the supercharge \eqref{Q-op} is also self-adjoint. Moreover, it follows from \eqref{Main-rel} that on the space of spherical harmonics, the eigenvalues $q_{j}$ of \eqref{Q-op} are of the form
\begin{equation*}
q_j = \pm (j + 1/2), \qquad j = 0, 1, 2,\ldots
\end{equation*}
Observe that while the angular momentum generators $J_1$, $J_2$, $J_3$ are symmetries of the quantum rotor $H = Q^2$, they are not symmetries of the supercharge $Q$; indeed, a direct calculation shows that $[J_i,Q]\neq 0$ for $i = 1, 2, 3$. It is also verified that the reflections $R_1$, $R_2$, $R_3$ do not commute with $Q$ either. 
\subsection{Symmetry algebra}
Let $K_1$, $K_2$, $K_3$ be the self-adjoint operators defined as
\begin{equation}
\label{realization}
K_1 = i J_1 R_2 + R_2 R_3/2, \qquad K_2 = -i J_2 R_1 R_2 + R_1 R_3/2,\qquad K_3 = -i J_3 R_1 + R_1 R_2/2.
\end{equation}
A straightforward calculation shows that these operators are symmetries of $Q$, i.e.
\begin{equation*}
[K_1, Q] = 0, \quad [K_2, Q] = 0, \quad [K_3, Q] = 0.
\end{equation*}
In view of \eqref{Main-rel}, it follows that $K_1$, $K_2$, $K_3$ are also symmetries of the quantum rotor. These operators are seen to obey the following relations:
\begin{equation}
\label{anti-spin}
\{K_1, K_2 \} = K_3, \qquad \{K_2, K_3\} = K_1, \qquad \{K_3, K_1\} = K_2.
\end{equation}
The relations \eqref{anti-spin} define $so(3)_{-1}$ or the anticommutator spin algebra. It is clear that these relations can be viewed as an anticommutator version of the relations \eqref{so3-comm}. The algebra \eqref{anti-spin} has a Casimir operator $C$ defined as
\begin{equation}
C = K_1^2 + K_2^2 + K_3^2,
\end{equation}
which commutes with $K_1$, $K_2$, $K_3$. This fact is easily confirmed using the elementary commutator identity $[A, BC]= \{A, B\} C - B \{A, C\}$. In the realization \eqref{realization}, the Casimir operator of the symmetry algebra is related to the supercharge $Q$ through
\begin{equation}
 C = Q^{2} - Q.
\end{equation}
The notation $so(3)_{-1}$ for the algebra \eqref{anti-spin} originates from the observation that the relations \eqref{anti-spin} correspond to taking $q\rightarrow -1$ in the defining relations of the quantum algebra $so(3)_q$, which read \cite{Fairlie-1990}
\begin{equation*}
q^{1/2} I_1 I_2 - q^{-1/2} I_2 I_1 = I_3,\qquad q^{1/2} I_2 I_3 - q^{-1/2} I_3 I_2 = I_1,\qquad q^{1/2} I_3 I_1 - q^{-1/2} I_1 I_3 = I_2.
\end{equation*}
Let us also note that the supercharge $Q$ defined in \eqref{Q-op} is not unique: there are other operators with similar expressions that square to the rigid rotor. For example, one could take 
\begin{equation*}
\widetilde{Q} = -i J_1 R_1 R_2 + i J_2 R_1 - i J_3 R_1 R_3 -R_1 R_2 R_3/2,
\end{equation*}
which also satisfies $\widetilde{Q}^2 = H$. Other operators, associated to different coproducts of the Lie superalgebra $osp(1,2)$ can also be constructed; see \cite{Genest-2014-2} for more details.
\section{Irreducible representations of $so(3)_{-1}$}
In this Section, we construct a basis for the irreducible representations of $so(3)_{-1}$ in terms of spherical harmonics, and compute the action of the generators on this basis. We proceed in two steps. First, we diagonalize the operator $K_3$ and obtain its eigenfunctions as linear combination of two spherical harmonics. Second, we diagonalize the supercharge $Q$ and obtain its eigenfunctions. The final expression for the common eigenfunctions of $K_3$ and $Q$ are expressed as a linear combination of four spherical harmonics.

 Let us first observe that the reflection operators have a simple action on the spherical harmonics $Y_{j}^{m}(\hat{r})$. Indeed, using the explicit expression \eqref{Yjm}, it is directly verified that
\begin{equation}
R_1 Y_{j}^{m}(\hat{r}) = Y_{j}^{-m}(\hat{r}),\quad R_2 Y_{j}^{m}(\hat{r}) = (-1)^{m} Y_{j}^{-m}(\hat{r}),\quad R_3 Y_{j}^{m}(\hat{r}) = (-1)^{j + m} Y_{j}^{m}(\hat{r}).
\end{equation}
\subsection{Diagonalization of $K_3$}
We now construct the eigenfunctions of the symmetry $K_3$ in the spherical harmonics basis. Using the above relations, as well as the formulas \eqref{J3} and \eqref{realization}, one finds that $K_3$ has the action
\begin{equation*}
K_3 Y_{j}^{m}(\hat{r}) = -i m Y_{j}^{-m}(\hat{r}) + \frac{(-1)^m}{2} Y_{j}^{m}(\hat{r}),\qquad m = -j, -j + 1, \ldots, 0, \ldots, j-1, j.
\end{equation*}
For a given $j$, the matrix representing $K_3$ in the spherical harmonics basis is thus block diagonal with $j$ $2\times 2$ blocks and one $1\times 1$ block, corresponding to $Y_{j}^{0}(\hat{r})$. This matrix is easily diagonalized. Let $M_{j}^{m, \epsilon}(\hat{r})$ be the functions defined as
\begin{equation}
\label{Aj}
M_{j}^{m,\epsilon}(\hat{r}) = \frac{1}{\sqrt{2}}\left[Y_{j}^{-m}(\hat{r}) + i \epsilon Y_{j}^{m}(\hat{r}) \right], \qquad \epsilon = \pm 1, \qquad m = 0, 1,\ldots, j.
\end{equation}
For $m = 0$, one should only consider the value $\epsilon = +1$. With this convention, the space spanned by the basis functions $M_{j}^{m,\pm}$ is $(j + 1)$-dimensional for $\epsilon = +1$ and $j$-dimensional for $\epsilon = -1$. The functions $M_{j}^{m,\epsilon}(\hat{r})$ satisfy the eigenvalue equation
\begin{equation*}
K_3 \,M_{j}^{m,\epsilon}(\hat{r}) = \left(\epsilon m + \frac{(-1)^m}{2} \right)\,M_{j}^{m,\epsilon}(\hat{r}).
\end{equation*}
Since the functions $M_{j}^{m,\epsilon}(\hat{r})$ are normalized linear combinations of the spherical harmonics, one has
\begin{equation*}
\int_{S^2} M_{j}^{m, \epsilon}(\hat{r}) [M_{j'}^{m', \epsilon'}(\hat{r})]^{*} \,\mathrm{d}\Omega = \delta_{jj'}\delta_{mm'}\delta_{\epsilon \epsilon'}.
\end{equation*}
The inverse relations are
\begin{equation*}
Y_{j}^{m}(\hat{r}) = -\frac{i}{\sqrt{2}}[M_{j}^{m, +}(\hat{r})- M_{j}^{m,-}(\hat{r})],\qquad Y_{j}^{-m}(\hat{r}) = \frac{1}{\sqrt{2}}[M_{j}^{m,+}(\hat{r}) + M_{j}^{m,-}(\hat{r})],\qquad m = 0, 1,\ldots, j.
\end{equation*}
It can be verified that the functions $M_{j}^{m, \epsilon}(\hat{r})$ are not eigenfunctions of the supercharge $Q$. In the next Subsection, we construct the eigenfunctions of $Q$ in the $M_{j}^{m, \epsilon}(\hat{r})$ basis.
\subsection{Simultaneous diagonalization of $Q$ and $K_3$}
In order to construct a basis in which the symmetry $K_3$ and the supercharge $Q$ are both diagonal, we examine the action of $Q$ on the eigenfunctions of $K_3$. Upon using the explicit expression \eqref{Q-op} for the supercharge $Q$, as well as the actions \eqref{Jpm-Action} of the raising/lowering operators and the explicit decomposition \eqref{Aj} of the eigenfunctions $M_{j}^{m,\epsilon}(\hat{r})$, a direct calculation shows that
\begin{multline}
Q M_{j}^{m,\epsilon}(\hat{r}) = \frac{(-1)^{j}i [(-1)^{m+1}-\epsilon]}{2}\;\sqrt{(j - m)(j + m +1)} \,M_{j}^{m + 1, \epsilon}(\hat{r})
\\
-\frac{1+ 2 (-1)^{m}\epsilon m}{2}\,M_{j}^{m,\epsilon}(\hat{r}) + \frac{(-1)^{j}i [\epsilon - (-1)^{m}]}{2}\;\sqrt{(j + m)(j - m +1)} \,M_{j}^{m - 1, \epsilon}(\hat{r}).
\end{multline}
As can be seen from the above expression, when $m$ is even and $\epsilon = +1$, or when $m$ is odd and $\epsilon = -1$, the coefficient in front of $M_{j}^{m - 1, \epsilon}(\hat{r})$ vanishes. Similarly, when $m$ is odd and $\epsilon = +1$, or when $m$ is even and $\epsilon = -1$, the coefficient in front of $M_{j}^{m + 1, \epsilon}(\hat{r})$ is zero. As a consequence, the matrix representing $Q$ can easily be diagonalized in the $M_{j}^{m, \epsilon}(\hat{r})$ basis. Let $\mathcal{F}_{j}^{k}(\hat{r})$ be defined as
\begin{equation}
\mathcal{F}_{j}^{k}(\hat{r}) = \sqrt{\frac{j-k}{2j +1}} M_{j}^{k+1,(-1)^{k}}(\hat{r}) + i (-1)^{j+k+1}\sqrt{\frac{j+k+1}{2j+1}} M_{j}^{k+1,(-1)^{k}}(\hat{r}),\qquad k = 0,1,\ldots, j.
\end{equation}
The functions $\mathcal{F}_{j}^{k}(\hat{r})$ satisfy
\begin{equation}
\label{Eigen-F}
Q \, \mathcal{F}_{j}^{k}(\hat{r}) = -(j+1/2)\,\mathcal{F}_{j}^{k}(\hat{r}),\qquad K_3\, \mathcal{F}_{j}^{k}(\hat{r}) = (-1)^{k}(k+1/2)\,\mathcal{F}_{j}^{k}(\hat{r}),
\end{equation}
and are thus simultaneous eigenfunctions of $Q$ and $K_3$. It is obvious that $\mathcal{F}_{j}^{k}(\hat{r})$ satisfies
\begin{align*}
\int_{S^2} \mathcal{F}_{j}^{k}(\hat{r}) [\mathcal{F}_{j'}^{k'}(\hat{r})]^{*} \,\mathrm{d}\Omega =\delta_{jj'}\delta_{kk'}.
\end{align*}
Similarly, let $\mathcal{G}_{j}^{k}(\hat{r})$ be another family of functions defined as
\begin{equation}
\mathcal{G}_{j}^{k}(\hat{r}) = \sqrt{\frac{j+k+1}{2j+1}} M_{j}^{k+1,(-1)^{k}}(\hat{r}) + i (-1)^{k+j}\sqrt{\frac{j-k}{2j+1}} M_{j}^{k,(-1)^{k}}(\hat{r}),\qquad k=0,\ldots, j-1.
\end{equation}
These functions $\mathcal{G}_{j}^{k}(\hat{r})$ satisfy
\begin{equation}
Q \, \mathcal{G}_{j}^{k}(\hat{r}) = (j+1/2)\,\mathcal{G}_{j}^{k}(\hat{r}),\qquad K_3\, \mathcal{G}_{j}^{k}(\hat{r}) = (-1)^{k}(k+1/2)\,\mathcal{G}_{j}^{k}(\hat{r}),
\end{equation}
and are again simultaneous eigenfunctions of $Q$ and $K_3$. They also obey the orthogonality relation
\begin{align*}
\int_{S^2} \mathcal{G}_{j}^{k}(\hat{r}) [\mathcal{G}_{j'}^{k'}(\hat{r})]^{*} \,\mathrm{d}\Omega =\delta_{jj'}\delta_{kk'}.
\end{align*}
We have thus constructed two orthonormal bases of joint eigenfunctions of the supercharge $Q$ and the symmetry operator $K_3$. For a given value of $j$, the set of eigenfunctions $\mathcal{F}_{j}^{k}(\hat{r})$ is $(j+1)$-dimensional and corresponds to the eigenvalue $-(j+1/2)$ of $Q$, while the set of eigenfunctions $\mathcal{G}_{j}^{k}(\hat{r})$ is $j$-dimensional and corresponds to the eigenvalue $(j+1/2)$ of $Q$. Since $Q$ is self-adjoint, the two bases are orthogonal to one another. It is also manifest that taken together they span the $(2j+1)$-dimensional space of spherical harmonics.
\subsection{Action of the symmetry algebra}
Let us now show that the span of $\mathcal{F}_{j}^{k}(\hat{r})$ for $k=0,\ldots, j$ supports an irreducible representation of the $so(3)_{-1}$ algebra. To that end, we examine the action of the generator $K_1$ on the basis elements $\mathcal{F}_{j}^{k}(\hat{r})$. By a direct calculation, one finds that $K_1$ acts in a tridiagonal fashion according to
\begin{align}
K_1 \, \mathcal{F}_{j}^{k}(\hat{r}) = U_{k+1}\,\mathcal{F}_{j}^{k+1}(\hat{r}) + B_{k} \mathcal{F}_{j}^{k}(\hat{r}) + U_{k} \mathcal{F}_{j}^{k-1}(\hat{r}),\qquad k=0,\ldots, j,
\end{align}
where
\begin{align}
\label{A}
U_{k} =
\begin{cases}
0 & k = 0
\\
\sqrt{\frac{(j+k+1)(j+1-k)}{4}} & k\neq 0
\end{cases},\qquad 
B_{k} =
\begin{cases}
(-1)^{j}\left(\frac{j+1}{2}\right) & k = 0
\\
0 & k\neq 0
\end{cases}.
\end{align}
The action of $K_2$ is directly obtained from the relation $K_2 = \{K_1, K_3\}$. Since $U_{k} \neq 0$ for $k = 1,\ldots, j$, it follows that the $so(3)_{-1}$ algebra acts irreducibly on the space spanned by the functions $\mathcal{F}_{j}^{k+1}(\hat{r})$. In other words, the functions $\mathcal{F}_{j}^{k}(\hat{r})$, for $k = 0,\ldots, j$, support a $(j+1)$-dimensional unitary irreducible representation of $so(3)_{-1}$. In a similar fashion, the span of $\mathcal{G}_{j}^{k}(\hat{r})$ for $k=0,\ldots, j-1$ also supports an irreducible representation of $so(3)_{-1}$. The action of $K_1$ on $\mathcal{G}_{j}^{k}(\hat{r})$ is given by
\begin{align}
K_1 \, \mathcal{G}_{j}^{k}(\hat{r}) = V_{k+1}\,\mathcal{G}_{j}^{k+1}(\hat{r}) + C_{k} \mathcal{G}_{j}^{k}(\hat{r}) + V_{k} \mathcal{G}_{j}^{k-1}(\hat{r}),\qquad k=0,\ldots, j-1,
\end{align}
where 
\begin{align}
\label{B}
V_{k} =
\begin{cases}
0 & k = 0
\\
\sqrt{\frac{(j+k)(j-k)}{4}} & k\neq 0
\end{cases},\qquad 
C_{k} =
\begin{cases}
(-1)^{j-1}\left(\frac{j}{2}\right) & k = 0
\\
0 & k\neq 0
\end{cases}.
\end{align}
By the same reasoning, the functions $\mathcal{G}_{j}^{k}(\hat{r})$, for $k=0,\ldots, j-1$, span a unitary irreducible representation of $so(3)_{-1}$ of dimension $j$. Upon comparing the fomulas \eqref{A} and \eqref{B}, it is seen that the two representations differ only by their respective dimensions.
\subsection{Spherical harmonics and representations of $so(3)_{-1}$}
In light of the above results, we can give the explicit decomposition of the irreducible representations of $so(3)$ in irreducible representations of $so(3)_{-1}$. Upon denoting by $V_j$ the $(2j+1)$-dimensional irreducible representations $so(3)$ defined by the matrix elements \eqref{J3}, \eqref{Jpm-Action}, and by $S_j$ the $(j+1)$-dimensional representations of $so(3)_{-1}$ defined by the matrix elements \eqref{Eigen-F}, \eqref{A}, one has
\begin{equation}
V_j = S_{j} \oplus S_{j-1}.
\end{equation}
In other words, each irreducible representation $V_{j}$ of $so(3)$ decomposes as a direct sum of two irreducible representations $S_{j}$ of $so(3)_{-1}$. A similar result was derived by Brown in \cite{Brown-2013} using different methods.
\section{Anti-Krawtchouk polynomials}
In this Section, we show that the anti-Krawtchouk polynomials arise as overlap coefficients between the basis $\mathcal{F}_{j}^{k}(\hat{r})$ and an alternative basis $\mathcal{Z}_{j}^{k}(\hat{r})$. For the remainder of this Section, we shall take $j = N$, where $N$ is a positive integer. 

Let us first recall that the basis $\mathcal{F}_{N}^{k}(\hat{r})$ with $k = 0,\ldots, N$ supports an  $(N+1)$-dimensional irreducible representation of the $so(3)_{-1}$ algebra and satisfies
\begin{align}
\label{Irrep-1}
\begin{aligned}
&\Gamma\, \mathcal{F}_{N}^{k}(\hat{r}) = -(N+1/2)\, \mathcal{F}_{N}^{k}(\hat{r}),
\\
& K_3 \,  \mathcal{F}_{N}^{k}(\hat{r}) = (-1)^{k}(k+1/2)\,  \mathcal{F}_{N}^{k}(\hat{r}),\qquad\qquad  k=0,1,\ldots, N,
\\
& K_1 \,  \mathcal{F}_{N}^{k}(\hat{r}) = U_{k+1}\,\mathcal{F}_{N}^{k+1}(\hat{r}) + B_{k} \mathcal{F}_{N}^{k}(\hat{r}) + U_{k} \mathcal{F}_{N}^{k-1}(\hat{r}),
\end{aligned}
\end{align}
with
\begin{align}
U_{k} =
\begin{cases}
0 & k = 0
\\
\sqrt{\frac{(N+k+1)(N+1-k)}{4}} & k\neq 0
\end{cases},\qquad 
B_{k} =
\begin{cases}
(-1)^{N}\left(\frac{N+1}{2}\right) & k = 0
\\
0 & k\neq 0
\end{cases}.
\end{align}
Let $\mathcal{Z}_{N}^{k}(\hat{r})$ be the basis functions defined by
\begin{align}
\label{Def-Z}
\mathcal{Z}_{N}^{k}(x_1, x_2, x_3) = \mathcal{F}_{N}^{k}(x_2, x_3, (-1)^{N+1}x_1),\qquad (x_1, x_2, x_3) = \hat{r} \in S^2.
\end{align}
It is verified that the basis $\mathcal{Z}_{N}^{k}(\hat{r})$ also supports an irreducible representation of $so(3)_{-1}$ in which the action of the generators is
\begin{align}
\label{Irrep-2}
\begin{aligned}
&\Gamma\, \mathcal{Z}_{N}^{k}(\hat{r}) = -(N+1/2)\, \mathcal{Z}_{N}^{k}(\hat{r}),
\\
& K_1 \,  \mathcal{Z}_{N}^{k}(\hat{r}) = (-1)^{k}(k+1/2)\,  \mathcal{Z}_{N}^{k}(\hat{r}),\qquad\qquad  k=0,1,\ldots, N,
\\
& K_2 \,  \mathcal{Z}_{N}^{k}(\hat{r}) = U_{k+1}\,\mathcal{Z}_{N}^{k+1}(\hat{r}) + B_{k} \mathcal{Z}_{N}^{k}(\hat{r}) + U_{k} \mathcal{Z}_{N}^{k-1}(\hat{r}).
\end{aligned}
\end{align}
Consider the expansion coefficients $W_{n}(k)$ between the bases $\mathcal{F}_{N}^{k}(\hat{r})$ and $\mathcal{Z}_{N}^{k}(\hat{r})$. These coefficients appear in the formula
\begin{align}
\label{Expansion-1}
\mathcal{Z}_{N}^{k}(\hat{r}) = \sum_{n=0}^{N} W_{n}(k)\, \mathcal{F}_{N}^{n}(\hat{r}),
\end{align}
which holds for any value of the position vector $\hat{r}\in S^2$. The overlap coefficients can be expressed in terms of the integral
\begin{align*}
W_{n}(k) = \int_{S^2} \mathrm{d}\Omega\; \mathcal{F}_{N}^{n}(\hat{r}) [\mathcal{Z}_{N}^{k}(\hat{r})]^*.
\end{align*}
Upon applying $K_1$ on both sides of the expansion formula \eqref{Expansion-1} and using the actions \eqref{Irrep-1} and \eqref{Irrep-2}, it is seen that the expansion coefficients $W_{n}(k)$ obey the three-term recurrence relation
\begin{align*}
(-1)^{k}(k+1/2)\; W_{n}(k) = U_{n+1} W_{n+1}(k) + B_{n} W_{n}(k) + U_{n} W_{n-1}(k), \qquad n=0,\ldots, N.
\end{align*}
Upon taking $W_{n}(k) = \omega_k \widehat{P}_{n}(y_k)$ with $\widehat{P}_{0}(y_k) = 1$ and
\begin{align}
\label{Weight}
w_{k} = \int_{S^2} \mathrm{d}\Omega\; \mathcal{F}_{N}^{0}(\hat{r}) [\mathcal{Z}_{N}^{k}(\hat{r})]^*,
\end{align}
one finds that the monic polynomials $\widehat{P}_{n}(y_k)$ satisfy the recurrence relation
\begin{align*}
y_k \widehat{P}_{n}(y_k) = U_{n+1}\widehat{P}_{n+1}(y_k) + B_{n} \widehat{P}_{n}(y_k) + U_{n} \widehat{P}_{n-1}(y_k),
\end{align*}
with $y_k=(-1)^{k}(k+1/2)$. If one takes $\widehat{P}_{n}(y_k) = \left(\frac{2^n}{U_1\cdots U_{n}}\right) P_{n}(y_k)$, one finds the relation
\begin{align}
\label{Recurrence-1}
x_k P_{n}(x_k) = P_{n+1}(x_k) - (A_n + C_n) P_{n}(x_k) + A_{n-1}C_{n} P_{n-1}(x_k),
\end{align}
where 
\begin{align}
\label{Recurrence-Coeff}
A_n =\frac{(-1)^{n+N+1}(N+1)+n+1}{4},\qquad 
C_{n} = 
\begin{cases}
0 & n=0
\\
\frac{(-1)^{N+n}(N+1)-n}{4} & n\neq 0
\end{cases},
\end{align}
and where
\begin{align}
\label{x}
x_k = (-1)^{k}(k/2+1/4)-1/4,\qquad k = 0, 1, \ldots, N.
\end{align}
The recurrence relation \eqref{Recurrence-1} corresponds to that of the Bannai-Ito polynomials $B_{n}(x_k; \rho_1, \rho_2, r_1, r_2)$, in the notation of \cite{Genest-2013-2}, with 
\begin{align}
\label{parameters}
\rho_1 = 0,\qquad \rho_2 =\frac{(-1)^{N}(N+1)}{2},\qquad r_1=0, \qquad r_2 = \frac{(-1)^{N}(N+1)}{2},
\end{align}
with $x_k$ given by \eqref{x}. As the monic polynomials $P_{n}(x_k)$ are associated with the anti-commutator spin algebra, we shall refer to them as the Anti-Krawtchouk polynomials. The name Krawtchouk is justified by the fact that when calculating the overlap coefficients between two spherical harmonics differing by a cyclic permutation of the coordinates, as in \eqref{Def-Z} (without the change of sign), one naturally finds the Krawtchouk polynomials with parameter $p=1/2$. The explicit hypergeometric expressions for the polynomials $P_{n}(x_k)$ involves a ${}_4F_3$ with argument $1$ whose explicit expression can be found in \cite{Genest-2013-2} using the choice of parameters \eqref{parameters}. Finally, the polynomials $P_{n}(x_k)$ satisfy a discrete orthogonality relation of the form
\begin{align*}
\sum_{k=0}^{N} \omega_{k}  P_{n}(x_k) P_{m}(x_k) = u_1\cdots u_{n} \delta_{nm},
\end{align*}
where the weight $w_{k}$ is given by (up to a normalization factor)
\begin{align*}
w_{k} = (-1)^{k}\times
\begin{cases}
\frac{(1+\alpha_N)_{k}}{(1-\alpha_N)_{k}} & k \text{ even}
\\
\frac{(1+\alpha_N)_{k-1}}{(1-\alpha_N)_{k-1}} & k \text{ odd}
\end{cases},
\end{align*}
where $\alpha_{N} = (-1)^{N}(N+1)$; thus result can be found in \cite{Genest-2014-2}.
\section*{Conclusion}
In this paper, we have exhibited the supersymmetric nature of the quantum rotor by explicitly constructing a supercharge operator $Q$ whose square is the Hamiltonian of the rotor. We have obtained the symmetries of $Q$, and shown that these symmetries generate the $so(3)_{-1}$ algebra, also known as the anticommutator spin algebra. We have constructed a basis for the eigenfunctions of the supercharge in terms of spherical harmonics, and shown that each $(2j+1)$-dimensional irreducible representation of $so(3)$ decomposes in a direct sum of two irreducible representations of $so(3)_{-1}$ of dimension $j$ and $(j+1)$. Finally, we have explained how a special family of orthogonal polynomials, the Anti-Krawtchouk polynomials, arise as overlap coefficients between two eigenbases of the supercharge corresponding to the diagonalization of two different symmetry operators.

These results call for a generalization to the multivariate case. Indeed, it would be interesting to determine a supercharge for the generalized rotor on the $n$-sphere, whose Hamiltonian is the $so(n)$ Casimir operator
\begin{align*}
H = \sum_{i < j\leq n } J_{ij}^2,
\end{align*}
with $J_{ij}=\frac{1}{i}(x_i \partial_{x_j}-x_{j} \partial_{x_i})$. In this case, one would expect the symmetry algebra of the supercharge to correspond to a special case of the higher rank Bannai--Ito algebra, as determined in \cite{DeBie-2015}.
\section*{Acknowledgments}
VXG holds a postdoctoral fellowship from the Natural Sciences and Engineering Research Council of Canada (NSERC). The research of LV is supported in part by NSERC. GFY is funded by  the National Natural Science Foundation of China (Grant \#11371251)
\small
\bibliographystyle{plain}
%\bibliography{ref}

\end{document}